\documentclass{webofc}
\usepackage[varg]{txfonts}   % Web of Conferences font
\usepackage{hyperref}
\usepackage{url}
\hypersetup{colorlinks=true,citecolor=blue,urlcolor=blue,linkcolor=blue}
\newcommand{\Rho}[1]{\rho_{\scalebox{.7}{#1}}}
\newcommand{\Cas}[1]{C_{\scalebox{.7}{#1}}}
\def\bm#1{\mbox{\boldmath$#1$}}
\newcommand{\Nc}{\ensuremath{N_{\rm c}}\xspace}
\newcommand{\CF}{C_{\scalebox{.6}{\rm F}}}
\newcommand{\dd}{{\rm d}}
\usepackage{tikz}

\usetikzlibrary{patterns}
\usetikzlibrary{decorations.pathmorphing,decorations.markings,trees,calc,shapes.geometric}
\tikzset{
  qua/.style={postaction={decorate},decoration={markings,mark=at position .55 with {\large\arrow{>}}}},
  anti/.style={postaction={decorate},decoration={markings,mark=at position .55 with {\large\arrow{<}}}},
  glu/.style={draw = black, decorate,decoration={coil,amplitude=1.4pt, segment length=2.6pt}} ,
  glu2/.style={draw = blue} ,
  vertical align/.style={baseline=-.5*(height("$+$")-depth("$+$"))},
  every picture/.style={vertical align}
}

\def \drawqua (#1) {
\draw[postaction={decorate},decoration={markings,mark=at position #1 with {\large\arrow{>}}}] 
}
\def \drawanti (#1) {
\draw[postaction={decorate},decoration={markings,mark=at position #1 with {\large\arrow{<}}}] 
}

\newcommand{\tikeq}[1]{\vcenter{\hbox{\begin{tikzpicture}#1\end{tikzpicture}}}}

\newcommand{\drawdash}{\draw[dash pattern=on 1pt off 1.3pt, line width=1.2]}

\def \OrthoBra (#1,#2) {
\tikeq{  \begin{scope}[scale=#1]
 \coordinate (O) at (0,0); 
	        \coordinate (A2) at (-.5,-.3) ;  
	        \coordinate (A1) at (-1.1,-.3) ; 
	        \coordinate (A3) at (-.5,.3) ;  
	        \coordinate (A4) at  (-1.1,.3) ;  
    \draw (0.15,-1.1) node {\scalebox{.8}{$\bar{1}$}};
    \draw (0.15,-0.6) node {\scalebox{.8}{$\bar{2}$}};
    \draw (0.15,0.7) node {\scalebox{.8}{$4$}};
    \draw (0.15,1.1) node {\scalebox{.8}{$3$}}; 
	        \drawdash  ($(A1)+(0,.2)$) -- (A1) -- ++ (0,-.5) arc (180:270:.3) -- ++ (0.8,0) ;
	        \drawdash  ($(A2)+(0,.2)$) -- (A2) -- ++ (0,-.2) arc (180:270:.2) -- ++ (.3,0.) ;
	        \drawdash  ($(A4)+(0,-.2)$) -- (A4) -- ++ (0,.5) arc (180:90:.3) -- ++ (0.8,0);
	        \drawdash ($(A3)+(0,-.2)$) -- (A3) -- ++ (0,.2) arc (180:90:.2) -- ++ (.3,0);
   \fill[white,draw=black] ($(O)+(-.8,0)$) circle (.42);  
  \draw ($(O)+(-.8,0)$) node {$#2$};
  \end{scope}
 }}
\def \Balphabeta(#1,#2) {
\tikeq{  \begin{scope}[scale=#1]
	        \coordinate (O) at (0,0); 
	        \coordinate (A2) at  ($(O)+(-.5,0)$) ;  
	        \coordinate (A1) at  ($(A2)+(-.6,0)$) ; 
	        \coordinate (A3) at  ($(A2)+(0,1)$); 
	        \coordinate (A4) at  ($(A1)+(0,1)$);  
	        \coordinate (B2) at  ($(O)-(-.5,0)$) ;  
	        \coordinate (B1) at  ($(B2)-(-.6,0)$) ;  
	        \coordinate (B3) at  ($(B2)+(0,1)$);  
	        \coordinate (B4) at  ($(B1)+(0,1)$);                
	        \drawdash  ($(A1)+(0,.2)$) -- (A1) -- ++ (0,-.5) arc (180:270:.3) -- ++ (1.6,0) arc (-90:0:.3) -- ($(B1)+(0,.2)$) ;
	        \drawdash  ($(A2)+(0,.2)$) -- (A2) -- ++ (0,-.2) arc (180:270:.2) -- ++ (.6,0.) arc (-90:-0:.2) -- ($(B2)+(0,.2)$);
	        \drawdash  ($(A4)+(0,-.2)$) -- (A4) -- ++ (0,.5) arc (180:90:.3) -- ++ (1.6,0) arc (90:0:.3) -- ($(B4)+(0,-.2)$);
	        \drawdash ($(A3)+(0,-.2)$) -- (A3) -- ++ (0,.2) arc (180:90:.2) -- ++ (.6,0) arc (90:0:.2) -- ($(B3)+(0,-.2)$);
   \fill[white,draw=black] ($(O)+(-.8,.5)$) circle (.42);  
  \draw ($(O)+(-.8,.5)$) node {$\alpha$};
    \fill[white,draw=black] ($(O)+(.8,.5)$) circle (.42);  
  \draw ($(O)+(.8,.5)$) node {$\beta$};
 \draw[glu] ($(O)+(0,-0.8)$) -- ($(O)+(0,#2)$) ; 
  \end{scope}
 }}
\def \Mqg { \tikeq{
		\coordinate (O) at (0,0);
		\draw[glu] ($(O)+(0,0)$) -- ++(0.5,.1) coordinate(A);
		\draw[glu] (A) -- ++(0.5,-.1);
		\draw[qua] ($(O)+(0,.7)$) -- ++(0.5,-.1)coordinate(B);
		\draw[qua] (B) -- ++(0.5,.1);
		\draw[glu] (A) --(B);
}
 - \xi \ 
 \tikeq{
		\coordinate (A) at (.3,.0);
		\coordinate (B) at (.7,.0);
    \draw[qua] ($(A)-(.3,-.35)$) -- (A);
    \draw[qua] (A) -- (B);
    \draw[qua] (B) -- ($(B)+(.3,.35)$) ;
    \draw[glu] ($(A)-(.3,.35)$) -- (A);
    \draw[glu] (B) -- ($(B)+(.3,-.35)$) ;
}}
\def \Mqgxitoone { \tikeq{
		\coordinate (O) at (0,0);
		\draw[glu] ($(O)+(0,0)$) -- ++(0.5,.1) coordinate(A);
		\draw[glu] (A) -- ++(0.5,-.1);
		\draw[qua] ($(O)+(0,.7)$) -- ++(0.5,-.1)coordinate(B);
		\draw[qua] (B) -- ++(0.5,.1);
		\draw[glu] (A) --(B);
}
 - 
 \tikeq{
		\coordinate (A) at (.3,.0);
		\coordinate (B) at (.7,.0);
    \draw[qua] ($(A)-(.3,-.35)$) -- (A);
    \draw[qua] (A) -- (B);
    \draw[qua] (B) -- ($(B)+(.3,.35)$) ;
    \draw[glu] ($(A)-(.3,.35)$) -- (A);
    \draw[glu] (B) -- ($(B)+(.3,-.35)$) ;
  }
  = 
 \tikeq{
		\coordinate (A) at (.3,.0);
		\coordinate (B) at (.7,.0);
    \draw[qua] ($(A)-(.3,-.35)$) -- (B);
    \draw[qua] (B) -- (A);
    \draw[qua] (A) -- ($(B)+(.3,.35)$) ;
    \draw[glu] ($(A)-(.3,.35)$) -- (A);
    \draw[glu] (B) -- ($(B)+(.3,-.35)$) ;
}}
\begin{document}
\title{Medium-induced coherent gluon radiation for $2\to 2$ processes with general kinematics}

\author{\firstname{Greg} \lastname{Jackson}\inst{1}\fnsep\thanks{\email{jackson@subatech.in2p3.fr}} \and
        \firstname{St\'ephane} \lastname{Peign\'e}\inst{1}\fnsep\thanks{\email{peigne@subatech.in2p3.fr}} \and
        \firstname{Kazuhiro} \lastname{Watanabe}\inst{2}\fnsep\thanks{\email{kazuhiro.watanabe.b8@tohoku.ac.jp} (speaker)}
        % etc.
}

\institute{
SUBATECH UMR 6457 (IMT Atlantique, Universit\'e de Nantes, IN2P3/CNRS), 4 rue Alfred Kastler, 44307 Nantes, France
\and
Department of Physics, Tohoku University, Sendai 980-8578, Japan
 }

\abstract{
  High-energy proton-nucleus (pA) collisions 
  offer valuable insight into the role
  of cold nuclear matter effects in hadron production. 
  In particular, multiple rescatterings of an incoming parton by the nuclear target 
  are known to induce the radiation of many soft gluons. 
  Hadron production is affected by fully coherent energy loss (FCEL), owing to those radiated gluons with a long formation time.
  Here we present a recently derived formula for the induced single soft gluon radiation 
  spectrum beyond leading logarithmic accuracy, 
  whose main features are demonstrated with the example of $q\, g \to q\, g$ scattering.
}
\maketitle

\vspace{-0.75cm}

\section{Introduction}\label{intro}

Heavy flavor production in high-energy nuclear collisions has been an intriguing probe 
of 
QCD medium interactions at different stages of the collision~\cite{Apolinario:2022vzg}.
Experimental data accumulated at 
collider facilities like the 
LHC and RHIC  
heralds 
a precision era for studying physics in high-energy nuclear collisions, 
addressing QCD dynamics at 
high temperature and `cold' nuclear matter effects, 
which were the main topics discussed in the conference. 

Proton-nucleus (pA) collisions
have provided 
valuable
opportunities to study various 
effects of cold nuclear matter on particle production. 
Compared to proton-proton (pp) collisions, 
partons participating 
in hard 
scatterings in pA collisions 
will likely undergo multiple interactions in nuclear targets. 
Such multiple rescatterings induce the emission of many soft gluons that could have 
a formation time $t_f\sim \omega/k_\perp^2$ with $\omega$ and $k_\perp$ 
the energy and transverse momentum of the radiated gluon, respectively, larger than $L$, the length traversed through the medium~\cite{Arleo:2010rb}.
This results in the modification of hadron production rates 
due to fully coherent energy loss (FCEL).

The parametric dependence 
of the corresponding average energy loss
is 
$\Delta E_{\rm FCEL}\sim E$ 
with 
$E$ being the energy of the incoming parton from the projectile.
Of particular importance is that medium-induced coherent energy loss 
overwhelms the QCD Landau-Pomeranchuk-Migdal (LPM) effect, valid for $t_f<L$,
when $E\to \infty$ 
because $\Delta E_{\rm LPM}/E\to 0$~\cite{Baier:1996sk,Baier:1996kr,Zakharov:1996fv,Zakharov:1997uu}.
Therefore, medium-induced FCEL has proven to be crucial in 
explaining heavy meson ($J/\psi$, $D$) 
and 
light hadron 
nuclear suppression in pA collisions in 
a wide range of collision energies 
from FNAL, RHIC, to LHC~\cite{Arleo:2012hn,Arleo:2020eia,Arleo:2020hat,Arleo:2021bpv,Arleo:2021krm}.
Understanding both the qualitative and quantitative role of FCEL in 
suppressing hadron production over a broad kinematic range 
is a crucial task. Addressing this requires rigorous calculations
of the medium-induced soft gluon radiation spectrum. 

We have recently derived new results for 
this spectrum 
(for all $2\to 2$ scatterings in QCD) 
beyond leading logarithmic accuracy, 
enhancing the predictive power of FCEL estimations 
in the phenomenology of hadron production 
in pA collisions~\cite{Jackson:2023adv}. 
The general formula is valid in the full kinematic range 
of the underlying scattering processes. 
It should provide an important baseline to evaluate 
more accurately FCEL effects on hadron production. 
This paper aims to review
the medium-induced soft gluon spectrum
and illustrate
how the new formula for $2 \to 2$ scatterings can be  
`matched' to the original ones obtained for $2\to 1$ forward processes.

\section{Recap of FCEL at LLA}\label{sec:fcel}

Let us begin with the induced soft gluon spectrum 
for $2\to 1$ scattering, 
such as $qg\to q$ and $gg\to g\,$. 
We shall consider forward scattering 
of an asymptotic fast parton 
(labelled 1) 
with large $E$, 
traversing the target medium in the target rest frame. 
The fast parton from the projectile acquires (due to multiple soft scatterings in the medium) 
an averaged momentum kick $l_\perp^{\,2}\propto L$
in addition to a single hard scattering (by parton labelled 2) 
which delivers a transverse momentum $q_\perp$ to the final parton 
(labelled 3, and `tagged' with a transverse mass $m_\perp=\sqrt{m^2+q_\perp^2}\,$). 
It then induces soft gluon emission far before and after the scatterings. 
The induced soft-gluon spectrum $\dd I/\dd \omega$ 
is obtained from 
an interference between 
initial state gluon radiation and 
final state gluon radiation. 
In the 
{\em leading-log} approximation (LLA)~\cite{Peigne:2014uha}, it reads
\begin{align}
  \label{2to1_LL}
\left.\omega\frac{\dd I}{\dd \omega}\right|_{2\to 1}
  \approx&\ \
  F_c\frac{\alpha_s}{\pi}\left[\,
    \ln\left(1+\frac{l_{A\perp}^2E^2}{\omega^2 m_\perp^2}\right)
    -\big(l_{A\perp}\to l_{p\perp}\big)
    \,\right]
    \; .
\end{align}
Here, $F_c$ is a color factor given by a combination of Casimir operators associated with the partons 
involved in the hard scattering. A general rule for $2\to1$ processes is: $F_c = C_1 + C_3 - C_2$, 
where $C_{1,2}$ are the Casimirs of the incoming partons and $C_3$ is that of the final-state parton. 
For example: In $gg \to g$, $F_c = N_c > 0$ (energy loss) and in $qg \to q$, $F_c = -1/N_c < 0$ (energy gain).

Next, we consider the induced soft gluon spectrum 
for $2\to 2$ partonic process
 in the LLA, 
which is relevant for dijet and open heavy flavor production. 
Allowing both final partons to have mass $m$,
we introduce the usual Mandelstam variables;
$s=(p_1+p_2)^2$,  $t=(p_1-p_3)^2$ etc.,
and define: 
$\xi \equiv (m^2-t)/s$ and 
$\bar \xi \equiv (m^2-u)/s = 1 - \xi$.
Then $\xi \in [0,1]$
and the dijet invariant mass in the final state is
$M_\xi^2=m_\perp^2/(\xi(1-\xi))\,$. 
In the LLA,
the radiated soft gluon 
sees only the global color of the final state
(the final state is viewed as a ``point-like dijet'' by the radiation)~\cite{Peigne:2014rka}, 
 and we have:
\begin{align}
\label{2to2_LL}
\left.\omega\frac{\dd I}{\dd \omega}\right|_{2\to 2}
  =\sum_\alpha~\rho_\alpha~F_\alpha~\frac{\alpha_s}{\pi}
  \left[\,
  \ln\left(1+\frac{l_{A\perp}^2E^2}{\omega^2M_\xi^2}\right)
  -\big(l_{A\perp}\to l_{p\perp}\big)
  \,\right] 
  \; ,
\end{align}
where $\alpha$ labels an irreducible representation (irrep) of the color state of the final parton pair, with 
associated probability $\rho_\alpha$ and Casimir $C_\alpha\,$.
Here $F_\alpha = C_1+C_\alpha-C_2$ with $C_\alpha$ the global color charge of the final parton pair
and $C_{1,2}$ the Casimir of the initial partons. 
Comparing with eq.~\eqref{2to1_LL}, we see that $m_\perp$ is 
generalized to $M_\xi$, which now depends on the scattering kinematics 
via $\xi\,$. The LLA above requires that $\xi \sim \bar \xi \sim 1/2\,$, 
because the logarithm in eq.~\eqref{2to2_LL} becomes small for $\xi \to 0$ or $1\,$.

Let us note that the spectrum \eqref{2to2_LL} 
arises from the logarithmic domain 
$\omega M_\xi /E \ll  k_\perp \ll  l_{A\perp}$ with $k_\perp$ 
the radiated gluon transverse momentum, 
allowing to interpret physically the LLA as the limit of {\it purely non-abelian} medium-induced radiation (see Ref.~\cite{Peigne:2014rka} for a discussion).
In the next section, we will explain how eq.~\eqref{2to2_LL} 
may be generalized beyond this limit, 
to the full kinematic range
$\xi \in [0,1]\,$.

\section{Medium-induced gluon spectrum beyond LLA}\label{sec:spectrum}

The induced spectrum beyond LLA can be cast into the following compact form~\cite{Jackson:2023adv}:
\begin{align}
  \label{spectrum-0}
  \frac{\dd I}{\dd x}
  \ =\ 
  \Phi_{\alpha\beta}~S(x)_{\beta\alpha}
  \ =\ 
  \mathrm{Tr}\left[\Phi\cdot S(x)\right]
  \; ,
\end{align}
where $x \equiv \omega/E$ and $\Phi$ represents a {\em color density matrix},  
quantifying the entanglement between color components of 
the $2\to 2$ partonic scattering amplitude, 
\begin{align}
\label{phi}
\Phi_{\alpha\beta}
  \ \equiv\ 
  \frac{\mathrm{Tr}_{\rm Dirac}(\nu_\alpha \nu_\beta^\ast)}{\mathrm{Tr}_{\rm color}\mathrm{Tr}_{\rm Dirac}|M|^2}
  \; .
\end{align}
Here, $M_{12\to 34}=\sum_\alpha \nu_\alpha \langle \alpha |$ 
is a generic $2\to 2$ amplitude with 
Dirac structure $\nu_\alpha$, 
and with a color structure $\langle\alpha|$ 
conveniently defined with the help of `birdtracks'~\cite{Keppeler:2017kwt,Peigne:2024srm}:
\begin{equation}
\label{s-basis}
\langle {\alpha} \vert = \frac{1}{\sqrt{K_\alpha}} \ \OrthoBra(0.6,\alpha)  \ , 
\end{equation}
where the blob $\alpha$ denotes a projection (or transition) operator 
corresponding to an $s$-channel irrep $\alpha$ of the $12 \to 34$ partonic scattering process.
The factor $1/\sqrt{K_\alpha}$ [with $K_\alpha = {\rm dim}(\alpha)$] 
ensures 
$\langle \alpha \vert \beta \rangle = \delta_{\alpha \beta}$.\footnote{%
  Note that $\vert{\alpha}\rangle$ is obtained from $\langle{\alpha}\vert$ by complex conjugation, 
  which corresponds pictorially to taking the mirror image and reversing quark arrows, as needed 
  for ${M}^{\,\ast}_{12\to 34}\,$ in \eqref{phi}.
}

Meanwhile, $S$ is a matrix associated to the soft radiation, namely
\begin{align}
S(x)_{\alpha\beta}
  \ \equiv\  
  \frac{\alpha_s}{\pi x}\left(\, 
  \mathcal{L}_{\xi} B_{\alpha\beta} + \mathcal{L}_{\bar{\xi}} \overline{B}_{\alpha\beta}
  \, \right) \, = \frac{\alpha_s}{\pi x}\left(\, 
    \frac{\mathcal{L}_\xi+\mathcal{L}_{\bar\xi}}{2}(B_+)_{\alpha\beta}
  + \frac{\mathcal{L}_\xi-\mathcal{L}_{\bar\xi}}{2}(B_-)_{\alpha\beta}
 \,\right)
 \label{eq:spectrum_BLLA}
\end{align}
where the large soft factor $\mathcal{L}_\xi\approx \ln\left(1+\frac{\xi^2l_{A\perp}^2}{x^2m_\perp^2}\right)-(l_{A\perp}\to l_{p\perp})$,
and $B_{\alpha\beta}$ represents soft color connections between the initial state gluon radiation
and 
the final state gluon radiation: 
\begin{equation} 
  B_{\alpha\beta} \ \equiv\  
  \frac{2}{\sqrt{K_\alpha K_\beta}} \ 
  \Balphabeta(0.6,1.4)  
  \ \ , \quad
  \bar B_{\alpha\beta} \ \equiv\   
   \frac{2}{\sqrt{K_\alpha K_\beta}} \ 
  \Balphabeta(0.6,1.8) 
  \ \ . \label{eq:Bbar}
\end{equation}
Note that for $\xi = 0$ and $\xi = 1$, only one of the factors ${\cal L}_\xi$ or ${\cal L}_{\bar \xi}$ survives and the spectrum is akin to the $2\to 1$ formula of eq.~\eqref{2to1_LL}. 
On the right side of eq.~\eqref{eq:spectrum_BLLA}, $B_+$ ($B_-$) 
is the diagonal (off-diagonal) color matrix part.
For $\xi\sim 1/2$, we obtain 
${\mathcal L}_\xi\sim {\mathcal L}_{\bar{\xi}}$ so that 
only the term proportional to $B_+ \equiv B + \bar B$ survives. 
Using color conservation, $T_1 + T_2 = T_\alpha = T_3 + T_4\,$,
\begin{equation}
\left( B_{+} \right)_{\alpha\beta} 
=  \langle \alpha \vert 2 \, T_1 \, T_\alpha \vert\, {\beta} \rangle
=  \langle {\alpha} \vert T_1^2 + T_\alpha^2 - T_2^2  \vert\, \beta \rangle
= (C_1+ C_\alpha - C_2) \, \delta_{\alpha\beta} \, .
\end{equation}
In that particular case, 
there is no change in the color state of the final parton pair. 
On the contrary, beyond LLA with $\xi\neq 1/2$, 
the induced soft gluon can change the color state of the parton pair.
Explicit formulae for both $\Phi$ and $S$ can be found in Ref.~\cite{Jackson:2023adv}.

As a trace, ${\dd I}/{\dd x}$ 
is independent of the color basis but can be diagonalized in some basis.
Specifically, the induced spectrum beyond LLA can be matched to the results in LLA 
as follows:
\begin{align}
&\left.\frac{\dd I}{\dd x}\right|_{\xi=\frac12}
  \ =\ 
  \frac{\alpha_s}{\pi x}\sum_{\alpha^s} \Phi_{\alpha^s\alpha^s}^s~(C_1+C_{\alpha^s}-C_2)~\mathcal{L}_{\xi = 1/2},
  \label{spectrum-symmetric}\\
&\left.\frac{\dd I}{\dd x}\right|_{\xi=0}
  \ =\ 
  \frac{\alpha_s}{\pi x}\sum_{\alpha^t} \Phi_{\alpha^t\alpha^t}^t~(C_1+C_3-C_{\alpha^t})~\mathcal{L}_{\bar{\xi} = 1},
  \label{xitozero-spectrum-t}\\
&\left.\frac{\dd I}{\dd x}\right|_{\xi=1}
  \ =\ 
  \frac{\alpha_s}{\pi x}\sum_{\alpha^u} \Phi_{\alpha^u\alpha^u}^u~(C_1+C_4-C_{\alpha^u})~\mathcal{L}_{\xi = 1},
  \label{xitoone-spectrum-u}
\end{align}
where the superscript $s$, $t$, or $u$ indicates the channel in which the sum over available irreps $\alpha$ is to be considered, 
and in which the matrix $\Phi$ is expressed.

\section{Illustration: quark-gluon scattering}\label{sec:demo}

Here, we will consider the case of  $qg\to qg$ scattering as an illustration, 
where in accordance with the definitions below eq.~\eqref{2to2_LL}, 
$\xi$ corresponds to the energy fraction of the final gluon in 
the nucleus rest frame.
The amplitude for this process reads
\begin{equation}
  M_{qg \to qg} \ = \ \mathcal{F} \, \left[ \ \Mqg \ \right] \, ,
  \label{M_qg_to_qg}
\end{equation}
where the coefficient $\mathcal{F}$ 
contains most of the Lorentz structure, 
and the diagrams represent the color.

In the preferred basis for each of the limiting cases 
$\xi=\frac{1}{2}$, $\xi = 0$, and $\xi=1$, 
the spectrum takes the form specified by 
eq.~\eqref{spectrum-symmetric}, 
eq.~\eqref{xitozero-spectrum-t} and 
eq.~\eqref{xitoone-spectrum-u}, 
where the corresponding available irreps are 
$\alpha^s = \{ \bm{3}, \bm{\bar 6}, \bm{15} \}$, 
$\alpha^t= \{ \bm{{1}}, \bm 8 \}$, and 
$\alpha^u= \{ \bm{3}, \bm{\bar 6}, \bm{15} \}$ in the $s$, $t$ or $u$-channel, respectively. 
To be specific,\footnote{%
  Let us recall the needed Casimirs:
   $\Cas{\bm{{3}}} = \CF\, ,
   \Cas{\bm{{\bar 6}}} = \CF+\Nc-1\, ,
   \Cas{\bm{{8}}} = \Nc\, ,
   \Cas{\bm{{15}}} = \CF+\Nc+1\,
   $.
  }
we have
\begin{align}
\label{spectrum-symmetric-qg-0}
& \left. \frac{\dd I}{\dd x} \right|_{\xi = \frac{1}{2}}
\ = \ \frac{\alpha_s}{\pi \, x} \, {\cal L}_{\xi=1/2}  \, 
  \big[ \, \Rho{\bm{{3}}} \, (2\CF-\Nc)
  + \Rho{\bm{{\bar 6}}} \, (2\CF-1)
  + \Rho{\bm{{15}}} \, (2\CF+1)
  \, \big] 
\  , \\ 
\label{xitozero-spectrum-t-qg-0}
& 
\left. \frac{\dd I}{\dd x} \right|_{\xi \to 0} \ = \ 
\frac{\alpha_s}{\pi \, x} \, {\cal L}_{\bar{\xi}=1} \, 
  \big[ \, \Rho{\bm{1}}^t \, (2\CF) 
  + \Rho{\bm{8}}^t \,  (2\CF-\Nc ) \, \big] 
\  , \\ 
\label{xitoone-spectrum-u-qg-0}
& \left. \frac{\dd I}{\dd x} \right|_{\xi \to 1} \ = \ 
\frac{\alpha_s}{\pi \, x} \, {\cal L}_{\xi=1} \,  
  \big[ \, \Rho{\bm{{3}}}^u \, \Nc
  + \Rho{\bm{{\bar 6}}}^u
  - \Rho{\bm{{15}}}^u
  \, \big]  
\  .
\end{align}
It is obvious from eq.~\eqref{M_qg_to_qg}, that 
for $\xi \to 0\,$ the $t$-channel exchange is forced 
into an octet, i.e. 
$\Rho{\bm{1}}^t=0$ and $\Rho{\bm{8}}^t=1\,$.
Using color conservation, \scalebox{.6}{$\Mqgxitoone$}, 
one observes that for $\xi \to 1\,$, the $u$-channel exchange 
must occur in the triplet, i.e. 
  $\Rho{\bm{{3}}}^u = 1$ and
  $\Rho{\bm{{\bar 6}}}^u=\Rho{\bm{{15}}}^u = 0 \,$.
For $\xi = \frac12\,$, the $s$-channel probabilities are 
$\Rho{\bm{3}} = \frac{ (\Nc^2+1)^2 }{ (\Nc^2-1)(5\Nc^2-1) }=\frac{25}{88}\,$, 
$\Rho{\bm{\bar 6}} = \frac{ 2 \Nc^2(\Nc-2) }{ (\Nc-1)(5\Nc^2-1) }=\frac{9}{44}\,$, and
$\Rho{\bm{15}} = \frac{ 2\Nc^2(\Nc+2) }{ (\Nc+1)(5\Nc^2-1) }=\frac{45}{88}\,$,
satisfying $\Rho{\bm{3}} + \Rho{\bm{\bar 6}} + \Rho{\bm{15}} =1$.

\begin{figure}[h]
\centering
\includegraphics[width=6cm,clip]{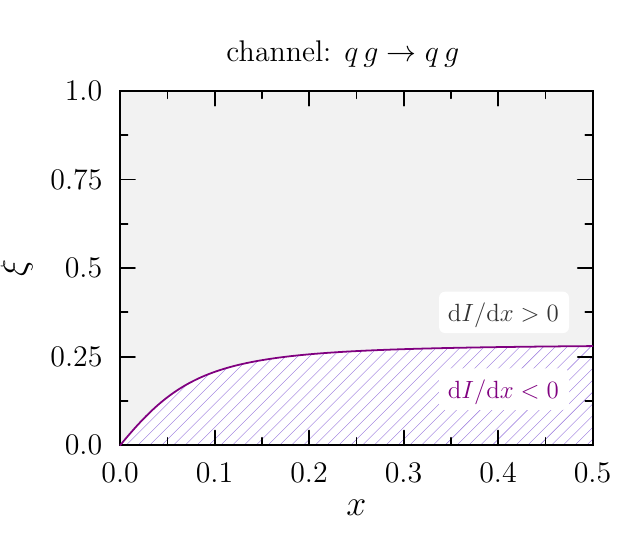}
\caption{
  Regions in the $(x,\xi)$-plane, corresponding 
  to energy-loss (solid gray) or energy-gain (hatched purple).
  In this figure, $\Nc=3$, $l_{\perp{\rm A}} =  \frac14 m_\perp$ and 
  $l_{\perp{\rm p}} =  \frac1{10} m_\perp\,$. 
  }
\label{fig-1}
\end{figure}

On the other hand, we may directly evaluate eq.~\eqref{spectrum-0} 
(using the matrices provided in Ref.~\cite{Jackson:2023adv}), 
to find, for all $\xi\,$, 
\begin{equation}
\label{qg-spec}
 \frac{\dd I}{\dd x} = 
 \frac{\alpha_s}{2\pi\, x} 
 \  \frac{1}{ \CF \xi^2 + \Nc \bar \xi }\  
 \bigg[ \,
    \Big( \Nc^2 - \xi^2 \Big) {\cal L}_{\xi}
    \, + \,  \left( \frac{\xi^2}{\Nc^2} - 2 \bar \xi \right) {\cal L}_{\bar \xi}
    \, \bigg] \ .
\end{equation}
From the above expression 
one can readily obtain the spectra for 
$\xi=\frac{1}{2}$, $\xi = 0$, and $\xi=1$: 
\begin{align}
\label{spectrum-symmetric-qg}
& \left. \frac{\dd I}{\dd x} \right|_{\xi = \frac{1}{2}} \ 
= \ \frac{\alpha_s}{\pi \, x} \, {\cal L}_{\xi={1/2}} \, 
  \frac{(\Nc^2-1)(4 \Nc^2 -1)}{\Nc(5\Nc^2-1)}
  \  , \\ 
\label{xitozero-spectrum-t-qg}
& 
\left. \frac{\dd I}{\dd x} \right|_{\xi \to 0} \ = 
\ \frac{\alpha_s}{\pi \, x} \, {\cal L}_{\bar{\xi}=1} \,  
  \bigg( -\frac{1}{\Nc} \bigg) \  , \\ 
\label{xitoone-spectrum-u-qg}
& \left. \frac{\dd I}{\dd x} \right|_{\xi \to 1} \ = \ 
\frac{\alpha_s}{\pi \, x} \, {\cal L}_{\xi=1} \,  
  \Nc
  \  .
\end{align}
We note that eq.~\eqref{xitozero-spectrum-t-qg} is negative, which 
may be attributed to fully coherent (medium-induced) energy {\em gain}. 
Indeed, it turns out that significant regions of the 
phase space exhibit a negative spectrum. 
This is demonstrated in Figure~\ref{fig-1}, 
which displays  a chart of the sign of ${\rm d}I/{\rm d}x$ 
in the $(x,\xi)$-plane, 
with a boundary indicating where the induced radiation spectrum vanishes.

We stress that such a negative {\em medium-induced} spectrum 
is not unphysical, owing to the 
prescription which defines ${\rm d}I/{\rm d}\omega\,$,
namely by a {\em difference} between the 
pA and pp situations. 
Be that as it may, for a projectile to incur less 
radiation within a nuclear target seems counter-intuitive. 
Let us appeal to the $2\to 1$ spectrum for insight. 
Indeed, the prefactor in eq.~\eqref{2to1_LL} contains a (strictly positive) 
part $\sim C_1+C_3$ of the total spectrum 
which originates from the abelian-like radiation.
The latter is focussed within a cone circumscribed by the 
scattering angle $\theta_s \equiv q_\perp/E\,$. 
Large-angle radiation $\theta > \theta_s$ is mainly associated to 
the non-abelian charge carried by parton 2 (which delivers $q_\perp$) 
and the corresponding part of the total spectrum is 
thus proportional to $C_2\,$. 
The total spectrum is an {\em increasing} function of $q_\perp\,$, but 
it would be incorrect to equate this with the additional soft 
rescatterings provided by the nucleus.

The medium-induced spectrum arises from an integration 
over the $k_\perp$ of the radiated gluon, or equivalently 
the angle $\theta \simeq k_\perp / \omega\,$. 
However, the abelian and non-abelian radiation cones are 
not affected in the same way by the presence of the medium.
Soft rescatterings do effectively grow $\theta_s\,$ 
(which increases the abelian radiation) 
but 
the non-abelian cone is ultimately capped at 
$\theta < q_\perp/\omega\,$, being fixed by the hard process and unchanged by the presence of the medium. 
Thus, the non-abelian component of the 
allocated radiation must actually shrink. 
In this way, we heuristically understand that $F_c \propto C_1+C_3-C_2$ 
in eq.~\eqref{2to1_LL} contains a positive part $C_1+C_3$ 
but also the negative part $-C_2\,$. 
Care is evidently needed when defining  
medium-induced observables.

\section{Summary}\label{sec:summary}

In order to disentangle parton energy-loss from other nuclear effects, 
such as nuclear parton distribution functions, 
it is important to control 
energy-loss 
calculations in a wide kinematic range.
For FCEL, this is particularly crucial for 
channels in forward open heavy flavor production 
in pA collisions. 
The effect should not be neglected when
extracting precise information about shadowing 
of the gluon distribution in target nuclei. 
Such studies, 
useful 
in a high-precision era with rich experimental programs, 
ought to include various observables 
like single hadron and dihadron/dijet production,
which could probe different energy-loss patterns. 
Building a quenching weight from the new medium-induced spectrum 
would be required 
in forthcoming phenomenological applications. 
We leave it for future work.

\subsection*{Acknowledgements}
This work is supported by JSPS KAKENHI Grant No. JP25K07286 and in part by the Agence Nationale de la Recherche (ANR) under grant ANR-22-CE31-0018.

%
% BibTeX or Biber users please use (the style is already called in the class, ensure that the "woc.bst" style is in your local directory)
% \bibliography{your_bib_file} % Replace "your_bib_file" with the actual name of your .bib file
%
% Non-BibTeX users please use
%

\end{document}